\documentclass[english,aps,pra,twocolumn,showpacs,superscriptaddress,floatfix]{revtex4-1}
\usepackage[T1]{fontenc}
\usepackage[latin1]{inputenc}
\usepackage{amsmath}
\usepackage{graphicx}
\usepackage{amssymb}
\usepackage{dcolumn} 
\usepackage{bm} 

\makeatletter

\usepackage{amsfonts}
\usepackage{epsfig}
\usepackage{dsfont}

\newcommand{\Eq}[1]{Eq.~(\ref{#1})}

\newcommand{\be}{\begin{equation}}
\newcommand{\bea}{\begin{eqnarray}}
\newcommand{\eea}{\end{eqnarray}}
\newcommand{\ee}{\end{equation}}

\newcommand{\Fig}[1]{Fig.~\ref{#1}}

\newcommand{\bra}[1]{\mbox{$\langle #1 |$}}
\newcommand{\ket}[1]{\mbox{$| #1 \rangle$}}

\newcommand{\quot}[1]{``#1''}

\newcommand{\E}{{\cal E}}
\newcommand{\C}{{\cal C}}
\newcommand{\F}{{\cal F}}

\def\Tr{\mbox{Tr}}

\usepackage{babel}
\makeatother

\begin{document}

\title{Ancilla-assisted sequential approximation of nonlocal unitary operations} 

\author{Hamed \surname {Saberi}}
\affiliation{Institute for Theoretical Physics, University of Regensburg, 93040 Regensburg, Germany}
\affiliation{Physics Department, Arnold Sommerfeld Center for Theoretical Physics, and Center for NanoScience, Ludwig-Maximilians-Universit\"{a}t, Theresienstr.~37, 80333 Munich, Germany}
\affiliation{Department of Physics, Shahid Beheshti University, G.C., Evin, Tehran 19839, Iran}

\date{August 18, 2011}

\begin{abstract}

  We consider the recently proposed ``no-go'' theorem of Lamata \emph{et al} [Phys. Rev. Lett. \textbf{101}, 180506 (2008)]
  on the impossibility of sequential implementation of global unitary operations with the aid of an itinerant ancillary system
  and view the claim within the language of Kraus representation. By virtue of an extremely useful tool for analyzing entanglement properties of quantum operations, namely, operator-Schmidt decomposition, we provide alternative proof to the no-go theorem and also study the role of initial correlations between the qubits and ancilla in sequential preparation of unitary entanglers. Despite
  the negative response from the no-go theorem, we demonstrate explicitly how the matrix-product operator (MPO) formalism provides a flexible
  structure to develop protocols for sequential implementation of such entanglers with an optimal fidelity. The proposed numerical technique, which we call variational matrix-product operator (VMPO), offers a computationally efficient tool for characterizing the ``globalness'' and entangling capabilities of nonlocal unitary operations.

\end{abstract}

 	

\maketitle

\section{Introduction}
\label{sec:intro}

\begin{figure}[b]
\centering
\includegraphics[width=1\linewidth]{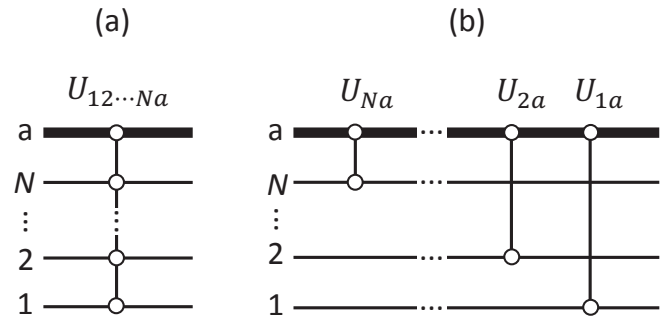}
\caption{Quantum circuit representation of (a) a generic nonlocal multiqubit unitary and (b) its ancilla-assisted sequential implementation. Each
bipartite unitary $U_{ka}$ acts \emph{only} on the Hilbert space of qubit $k$ and ancilla $a$ and leaves other qubits unchanged. To emphasize
this fact, instead of the common circuit representation of an arbitrary bipartite unitary by a rectangular box, we have used vertical lines with open circles at each qubit they act upon. Ancilla states are shown with solid bold lines throughout.}
\label{seq_factory}
\end{figure}

Engineering arbitrary global unitary operations entangling simultaneously multiple qubits is generically regarded to be a task of formidable difficulty, as it may require a complicated combination of exponentially many gates to approximate~\cite{Nielsen2000}. However, efficient implementation of such \emph{genuinely entangling} multiqubit operations is of paramount importance in quantum computation.
Recently, it has been wondered if the situation could
be facilitated by devising an ancilla-assisted decomposition of a multiqubit unitary operation into a sequence of ancilla-qubit unitary operations where each qubit is allowed to interact locally and only once with an itinerant ancillary system (e.g., a trapped multilevel atom coupled to a single mode of an optical cavity in the realm of cavity QED experiments~\cite{Schoen2005,Schoen2007,Saberi2009})
and without measurements~\cite{Lamata2008} (see \Fig{seq_factory}). Equivalently, according to the suggested scenario, instead of entangling directly neighboring qubits $k$ and $k+1$, an ancilla $a$ intervenes to get entangled with qubit $k$ and the ancilla state is swapped afterward with that of qubit $k+1$. Such decomposition, however, immediately faced a resounding no and was proved to be impossible in general for genuinely entangling unitaries, with the controlled-NOT (CNOT) gate being the handiest counterexample~\cite{Lamata2008}. A \emph{reductio ad absurdum} proof strategy for such a \quot{no-go} theorem implied the incompatibility of unitarity of the last constituent two-qubit operation and the deterministic nature of
the protocol. More precisely, having introduced an ancillary system to convey entanglement throughout the register of qubits, we eventually wish the ancilla \emph{decouples} from the qubits in the very last step. But it turns out that for arbitrary entangling unitary the ancilla always remains entangled with the qubit system, thereby spoiling the promised determinism.

In the present paper, we reconsider the issue of such a sequential quantum \quot{factory} of unitary entanglers within the framework of
Kraus representation~\cite{Kraus1983} and operator-Schmidt decomposition~\cite{Nielsen1998,Nielsen2003} and provide alternative proofs to the no-go theorem forbidding the realization of such a setup.
It turns out that employing such tools provides a systematic and transparent approach to various aspects of the problem and reveals the insurmountable mathematical obstacles to realization of such a scenario. Moreover, despite this negative answer, we investigate the
possibility of
implementing a nonlocal unitary with a certain imperfect \emph{fidelity}. In other words, we demonstrate it is always possible to satisfy both conditions of the sequentiality and unitarity of the ancilla-qubit operations at the same time, however, at the price of ending up with a sequentially implemented \emph{unitary} whose action is not perfectly equivalent to the original global unitary but is rather closest to that in some sense. This will be realized by exploiting the tools from matrix-product operator (MPO) theory~\cite{McCulloch2007,Crosswhite2008,Pirvu2010}. We also
study the role of initial correlations between the ancilla and qubits upon sequential preparation of unitaries and the way they could ever affect the
evolution of the joint ancilla-qubit system.

\section{Sequential quantum \quot{factory} of unitary entanglers viewed in the language of quantum operations}
\label{sec:Kraus}

\begin{figure}[t]
\centering
\includegraphics[width=0.83\linewidth]{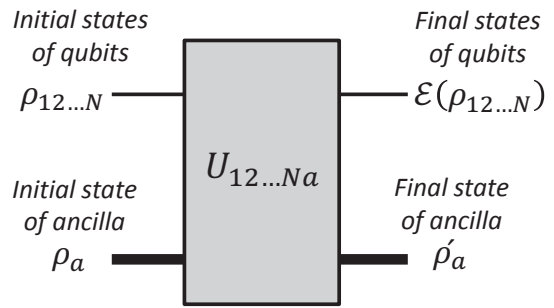}
\caption{Ancilla-assisted implementation of entanglers viewed as a quantum operation $\E(\rho_{12 \dots N})$. The joint ancilla-qubit system
evolves under a unitary transformation $U_{12 \dots Na}$.}
\label{quan_op}
\end{figure}

Let us first illustrate explicitly what is exactly meant by claiming that no entangling unitary can be implemented in a sequential way:
For this purpose, we suggest viewing the problem within the framework of quantum operations~\cite{Kraus1983,Nielsen2000} where the qubit
chain is regarded as the \emph{principal system} and the ancilla as the \emph{environment} (see \Fig{quan_op}). For simplicity, let
us consider for now the specific case of only two qubits, $N=2$. We shall later generalize our argument to the case of an arbitrary number of qubits $N$. Assuming that the qubits and ancilla start out in a product state $\rho_{12a}= \rho_{12} \otimes \rho_{\rm{a}}$, the evolution of the state of qubits is given by
\begin{eqnarray}
\label{eq:quantum_operation}
\hspace{-5mm} \E(\rho_{12})=\Tr_{\rm{a}}\bigl[ U_{12a} (\rho_{12} \otimes \rho_{\rm{a}}) U^{\dagger}_{12a}\bigr]= \sum_{k} E_k \rho_{12} E_k^{\dagger} \; ,
\end{eqnarray}
where $E_k \equiv {}_a\langle k | U_{12a} | 0\rangle_a$ are Kraus operators in an \emph{operator-sum representation}~\cite{Nielsen2000} of quantum operation $\E$ and
$|k\rangle_a$ represent an orthonormal basis for the
(finite-dimensional) state space of the ancilla; in particular,
$\ket{0}_a$ is assumed to be the initial state of the ancilla. We note that the aim is here to implement an entangling \emph{target} two-qubit unitary $U_{12}$ with the aid of an ancilla $a$ whose action should be eventually factorized as $U_{12a}= U_{12} \otimes \mathds{1}_a$. Should the latter be the case, the Kraus operators are simply given by
\begin{eqnarray}
\label{eq:Kraus_operators_anc_dec}
E_k= {}_a\langle k| U_{12} \otimes {\mathds{1}}_a |0\rangle_a= U_{12} \delta_{k0}  \; .
\end{eqnarray}
The sequential preparation of the two-qubit unitary $U_{12}$, on the other hand, suggests Kraus operators of the form
\begin{eqnarray}
\label{eq:Kraus_operators_seq}
E_k = {}_a\langle k | (U_{2a} \otimes \mathds{1}_{1}) (\mathds{1}_{2} \otimes U_{1a}) |0\rangle_a   \; .
\end{eqnarray}
Note that at step $k$ of the sequential implementation of a global $N$-body unitary operation $U_{12...N}$,
the two-body unitary $U_{ka}$ entangles only the ancilla $a$ and qubit $k$ and leaves other
qubits \emph{unchanged}, the latter action having been denoted by a properly indexed identity operator $\mathds{1}$ above.
Now, if the entangling unitary $U_{12}$ could be implemented sequentially, $E_1$ in \Eq{eq:Kraus_operators_seq} should also vanish.
Acting $E_1$ on some initial state of qubits 12 and performing Schmidt decomposition on the partition $1|a$ of the \emph{entangled} state $U_{1a}\ket{00}_{1a}$ yield
\begin{eqnarray}
\nonumber
E_1 \ket{0i}_{12}  & = & {}_a\langle 1 | (U_{2a} \otimes \mathds{1}_{1}) (\mathds{1}_{2} \otimes U_{1a}) |0i0\rangle_{12a}
\\
  \label{eq:vanishing_E0}
 & = &
\sum_{j=0}^{\chi}  {}_a\langle 1 | U_{2a}  |ij\rangle_{2a} \ket{\phi_j}_1= 0 \; ,
\end{eqnarray}
with the Schmidt rank $\chi \ge 1$~\cite{luc_proof}, and by convention we have incorporated the Schmidt coefficients into the Schmidt vectors~\cite{sch1}.
The orthogonality (and linear independence) of $\ket{\phi_j}_1$ implies that the only way for
$E_1$ to vanish for a sequentially prepared unitary would be that ${}_a\langle 1 | U_{2a}  |ij\rangle_{2a}$ vanish for all $i$ and
$j= 0, 1,\dots \hspace{1mm}$. The latter, however, causes two rows of $U_{2a}$ and, consequently, its determinant to vanish, thereby violating the invertibility and,
in turn, unitarity of $U_{2a}$.

For sequential implementation of a global unitary one should satisfy two \emph{constraints} of sequentiality and unitarity of the constituent ancilla-qubit operations. However, as illustrated above, these two cannot be fulfilled at the same time.

Note that the result above may be
readily extended to an arbitrary number of qubits $N$ since it is always possible to map this general case to that of only two unitaries $U_{2a} U_{1a}$ by combining the first $n-1$ unitaries and the $n$th one, where $n$ is the smallest step at which the ancilla decouples from the qubit chain~\cite{Lamata2008}. Therefore, for the sake of brevity, we shall henceforth only consider the paradigmatic case of $N=2$ and keep in mind that the results will accordingly be valid for an arbitrary number of qubits $N$ too.

We point out the \quot{no-go} statement above holds even if we relax the assumption that the ancilla and qubits start out in a product state.
To see this, we exploit the Fano form of the density matrix~\cite{Fano1983,Bengtsson2006} to follow the evolution of the joint ancilla-qubit system in the presence of initial correlations between the qubits as the \emph{open system}~\cite{Breuer2007} and the ancilla as the environment. Assuming the
action of the ancilla can be factorized like $U_{12a}= U_{12} \otimes \mathds{1}_a$, the quantum operation then is given by~\cite{Stelmachovic2001}
\begin{eqnarray}
\nonumber
\E(\rho_{12}) & = & U_{12} \rho_{12} U^{\dagger}_{12}
\\
\label{eq:init_corr_anc_dec_1}
&&  +
\sum_{k} {}_a\bra{k} U_{12a} \gamma_{ij}^{12a} \sigma_i^{12} \otimes \tau_j^a U^{\dagger}_{12a} \ket{k}_a   \; ,
\end{eqnarray}
where $\sigma_i$ and $\tau_j$ are the generators of SU(4) and SU($D$), respectively, and $D$ is the dimension of the Hilbert space of the ancilla. Straightforward algebra then yields
\begin{eqnarray}
\label{eq:init_corr_anc_dec_2}
\hspace{-4mm} \E(\rho_{12}) =
U_{12} \rho_{12} U^{\dagger}_{12} + \sum_{i,j} U_{12} \sigma_i^{12} U^{\dagger}_{12} \gamma_{ij}^{12a} \Tr\{ \tau_j^a \} \; .
\end{eqnarray}
But generators of SU($D$) are known to be traceless. So we conclude that for the case where the evolution of the whole ancilla-qubits is factorable with respect to the ancilla (or the ancilla decouples from the qubit chain), the initial correlations would not affect the final states of the qubits. Contrarily, for a nonfactorable sequentially implemented unitary, the correlation term in \Eq{eq:init_corr_anc_dec_2} does not necessarily
vanish, and initial correlations may play a significant role in the evolution of the joint ancilla-qubit system and, in turn, in the final states of both qubits and the ancilla.

The main obstacle in sequential implementation of an entangling unitary, namely, the fact that the ancilla cannot be set to decouple from the qubit
chain in the last step, may also be expressed in terms of the well-known problem of \emph{separability criteria} in quantum information~\cite{Peres1996,Duer1999,Lewenstein2000,Guehne2010}: When we start from a 1-qubit biseparable state~\cite{bisep} of ancilla and
qubits $\rho_{12a}=\rho_{12} \otimes \rho_a$, straightforward algebra employing Kraus representation for both the ancilla and qubits implies that a factorable evolution $U_{12a}= U_{12} \otimes U_a$ leaves such separability invariant, i.e., $\rho'_{12a}= \rho'_{12} \otimes \rho'_a$ where a prime indicates the final states. However, this ceases to be the case for the ancilla-assisted sequential evolution considered above.

\section{Alternative proof to the no-go theorem based on operator-Schmidt decomposition}
\label{sec:op_Schmidt}

In this section we provide an alternative proof to the no-go theorem by virtue of an extremely useful and elegant decomposition known in operator algebra as \emph{operator-Schmidt decomposition}~\cite{Nielsen1998,Nielsen2003} according to which the bipartite ancilla-assisted unitaries may be decomposed as
\begin{subequations}
  \label{eq:op_Schmidt_U_ia}
\begin{eqnarray}
U_{2a}= \sum_{i=1}^{\chi_2} A_i^2 \otimes B_i^a  ,
\\
U_{1a}= \sum_{i=1}^{\chi_1} A_i^1 \otimes C_i^a \; ,
\end{eqnarray}
\end{subequations}
with $\{A_i^2\}, \{A_i^1\}, \{B_i^a\}$, and $\{C_i^a\}$ being some orthogonal (but not necessarily normalized) operator bases defined on
the Hilbert-Schmidt space of operators acting on qubits 2 and 1 and ancilla $a$, respectively. Of particular importance is $\chi_2$ ($\chi_1$), called the \emph{Schmidt number} of operator $U_{2a}$ ($U_{1a}$), defined as the number of nonzero coefficients in its operator-Schmidt decomposition. Applying the decomposition to each bipartite entangling unitary $U_{ka}$ yields
\begin{eqnarray}
\nonumber
U_{12a} & = &(U_{2a} \otimes \mathds{1}_{1}) (\mathds{1}_{2} \otimes U_{1a})
\\
\label{eq:op_Schmidt_seq}
& = &
\nonumber
(\sum_{i} A_i^2 \otimes B_i^a \otimes \mathds{1}_1) (\mathds{1}_{2} \otimes \sum_{j} A_j^1 \otimes C_j^a)
\\
& = &
\sum_{i,j} A_j^1 \otimes A_i^2 \otimes (B_i^a C_j^a)  \; .
\end{eqnarray}
Without loss of generality we may take $\chi_2= 2$; the same argument would hold for an entangling unitary of Schmidt number $\chi_2= 4$~\cite{note}.
In order that the ancilla decouples \emph{unitarily} from the qubit chain, it should be possible
for it to be factorized like $B_i^a C_j^a= U_a$ for all $i$ and $j$ and with $U_a$ some unitary operator on ancilla space.
This in particular implies that $B_1^a C_1^a= B_2^a C_1^a$, eventually suggesting $U_{2a}=(\sum_i A_i^2) \otimes B^a$ to be a nonentangling
operation. The latter, however, is in contradiction to the original assumption that $U_{2a}$ has a Schmidt number larger than 1, and this readily completes the alternative proof.

\section{Optimal implementation of nonlocal unitaries within the framework of matrix-product operators}
\label{sec:MPO}

Having accepted the impossibility of implementing a nonlocal unitary in a sequential way, we may wonder if we could rather try to approach
a \emph{target} unitary entangler within a sequential procedure, albeit in an approximate way and with an optimal fidelity. More precisely, the aim is then to find bipartite
\emph{unitaries} $U_{1a}$ and $U_{2a}$ that, when applied sequentially instead of the global unitary $U_{12a}$, minimize
the Frobenius norm distance~\cite{frob}
\begin{eqnarray}
\nonumber
\C &=& \Bigl \| U_{12a}-(U_{2a} \otimes \mathds{1}_1) (\mathds{1}_2 \otimes U_{1a}) \Bigr \|^2_{F}
\\
\label{eq:cost_func}
& = &
2D-2{\rm Re} \Bigl\lbrace \Tr \bigl[U^\dagger_{12a} (U_{2a} \otimes \mathds{1}_1) (\mathds{1}_2 \otimes U_{1a}) \bigr] \Bigr \rbrace   \; .
\end{eqnarray}
Note here that
we hereby require both sequentiality and unitarity conditions to be fulfilled without violating the above-mentioned no-go theorem:
In fact, these two plus the implicit condition that the sequentially implemented unitary should produce the same action of the global one
(i.e., the identity $U_{12a} \ket{\Psi} = (U_{2a} \otimes \mathds{1}_1) (\mathds{1}_2 \otimes U_{1a}) \ket{\Psi}$ for arbitrary state $\ket{\Psi}$) do save the no-go theorem and make such a distance never vanish. Without such a condition, we could, of course, freely choose the sequential unitaries so that the ancilla decouples unitarily in the last step.

\begin{figure}[t]
\centering
\includegraphics[width=0.8\linewidth]{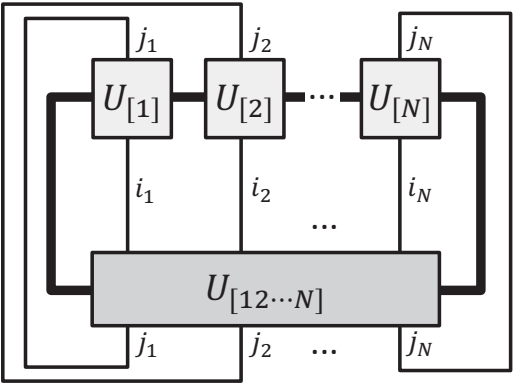}
\caption[MPO representation of sequential unitaries]{Graphical representation of the overlap of MPOs used to calculate the contractions in the cost function \Eq{eq:cost_MPO}. The square boxes represent $U_{[ka]}$ matrices of a sequentially prepared unitary, and the rectangular box represents the target \emph{global} unitary. The links connecting the boxes represent indices that are being contracted (or summed over).}
\label{opt_pat_op}
\end{figure}

The constrained nonlinear optimization problem~\cite{Bertsekas1996} associated with the cost function above may be solved efficiently by a variational search in the space of the MPOs arising from the sequential nature of the protocol. To see how operators of matrix-product form~\cite{McCulloch2007,Crosswhite2008,Pirvu2010} arise naturally within such a prescription, we express each ancilla-qubit unitary in terms of an orthonormal basis
\begin{eqnarray}
\label{eq:U_ia_basis}
U_{ka}=\sum_{i_k,j_k=0}^1 \sum_{\alpha_k,\beta_k=1}^D U_{j_k,\beta_k}^{i_k,\alpha_k} |i_k \alpha_k \rangle
\langle j_k \beta_k |  \; ,
\end{eqnarray}
where Roman (Greek) letters denote qubit (ancilla) indices. The sequentially implemented unitary, as a result, reads
\begin{widetext}
\begin{equation}
(U_{2a} \otimes \mathds{1}_1) (\mathds{1}_2 \otimes U_{1a}) =
\sum_{i_2, j_2} \sum_{\alpha_2, \beta_2} \sum_{i_1, j_1} \sum_{\alpha_1, \beta_1} U_{j_2,\beta_2 }^{i_2,\alpha_2} U_{j_1,\beta_1 }^{i_1,\alpha_1}
\label{eq:MPO}
|i_2 \alpha_2 \rangle \langle j_2 \beta_2 | \otimes |i_1 \alpha_1 \rangle \langle j_1 \beta_1|
= \sum_{i_2, j_2} \sum_{i_1, j_1} U_{[2]}^{i_2,j_2} U_{[1]}^{i_1,j_1}
|i_2 \rangle \langle j_2| \otimes |i_1 \rangle \langle j_1|  \; ,
\end{equation}
\end{widetext}
where we have defined in the last line $\sum_{\alpha_k \beta_k} U_{j_k,\beta_k}^{i_k,\alpha_k}
|\alpha_k\rangle \langle\beta_k| \equiv U_{[k]}^{i_k,j_k}$ to cast the original expression into products of unitary operators. We may now easily recognize (\ref{eq:MPO}) as an MPO already encountered in the calculation of finite-temperature density matrices~\cite{Verstraete2004}. With this, the cost function~(\ref{eq:cost_func}) in the MPO representation reads
\begin{eqnarray}
\nonumber
\hspace{-25mm} \C & =&  2D
\\
\label{eq:cost_MPO}
&& -
2{\rm Re}\Bigl\lbrace {\rm Tr}\biggl(\sum_{j_2',j_1'} \sum_{i_2,j_2} \sum_{i_1,j_1}
{U^{i_2 i_1, j_2' j_1'}_{[21]}}^\dagger U_{[2]}^{i_2,j_2} U_{[1]}^{i_1,j_1}
\nonumber
\\
&& \otimes |j_2' j_1'\rangle \langle j_2 j_1| \biggr) \Bigr\rbrace \; ,
\end{eqnarray}
where we have expressed the target unitary too, in terms of an orthonormal basis, and have accordingly defined
\begin{eqnarray}
\label{eq:Target_sum_anc_1}
\sum_{\alpha', \beta'} U_{\alpha' \beta'}^{i_2' i_1', j_2' j_1'}
\ket{\alpha'} \bra{\beta'} \equiv U_{[21]}^{i_2' i_1',j_2' j_1'}   \; 
\end{eqnarray}
to have
\begin{eqnarray}
\label{eq:Target_sum_anc_2}
U_{12a}= \sum_{i_2', j_2'} \sum_{i_1', j_1'} U_{[21]}^{i_2' i_1',j_2' j_1'} |i_2' i_1'\rangle \langle j_2' j_1'|   \; .
\end{eqnarray}
The calculation of the cumbersome summations on the right-hand side of Eq.~(\ref{eq:cost_MPO}) finds a very simple form in MPO representation
as illustrated in Fig.~\ref{opt_pat_op}. The minimization problem can be done very efficiently in an iterative
variational optimization technique (similar to those already used in the context of matrix-product states~\cite{Verstraete2004_2,Saberi2008,Saberi2009,Saberi2009_2,Weichselbaum2009})
in which we fix all but one of the sequential unitaries, let's say $U_{[k]}$, and minimize the cost function
\Eq{eq:cost_MPO} by varying over the matrix elements of $U_{[k]}$. In the next iteration, the neighboring unitary
is optimized, and once all the unitaries so obtained have been optimized locally, we sweep back again and so on until convergence.
In that sense, we suggest an extension of the previously developed method of variational matrix-product state (VMPS)~\cite{Saberi2008,Saberi2009_2,Weichselbaum2009} from states
to operators and suggest calling it the variational matrix-product operator (VMPO) method.

In the case that the ancilla is a qubit, i.e., $D=2$, the two-qubit sequential unitaries of the variational MPO in \Eq{eq:MPO} may be expanded in terms of the complete basis of Pauli matrices:
\begin{eqnarray}
\label{eq:Pauli_exp}
U_{[k]}=\exp(-i\sum_{j_1,j_2=0}^3 h_{j_1,j_2}^{[k]} \sigma_{j_1} \otimes \sigma_{j_2})   \; ,
\end{eqnarray}
where $h_{j_1,j_2}^{[k]}$ are real-valued coefficients and $\sigma_1$, $\sigma_2$, and $\sigma_3$ are the usual Pauli sigma matrices, with $\sigma_0\equiv \mathds{1}$ being the identity matrix.
The minimization of the cost function \Eq{eq:cost_MPO} then amounts to finding the optimal coupling matrix $h^{[k]}$ at
each step of the iterative optimization. The \emph{fidelity} of the sequential implementation of the global unitary $U_{12a}$ may be
quantified then as $\F=1-\tilde{\C}$, where $\tilde{\C}$ denotes the normalized converged cost function $\C$ of \Eq{eq:cost_MPO}
upon the very last sweep. The normalization is taken care of after dividing by the Frobenius norm of the involved operators.

We have applied the outlined procedure to some paradigmatic gates of quantum computing when the ancilla has the
dimension $D=2$, as illustrated in Table~\ref{table:F_D_2}: For instance, the two locally equivalent gates controlled-NOT ($\rm {CNOT}$) and controlled-$Z$ (CZ) can be implemented sequentially with a fidelity of 70.71\%. The 29.29\% error we associate with the existence of a ``fidelity gap,'' a strict theoretical \quot{red line} beyond which the global unitary cannot be implemented in a sequential fashion.

Similar optimization techniques may be exploited for higher ancilla dimension $D=4$, however, upon using the generators of SU(4) for the ancilla instead of Pauli matrices in \Eq{eq:Pauli_exp} to see if it has any influence on the values of the gaps. We have performed such a simulation for the same gates as in Table~\ref{table:F_D_2} and have found that the values of the gaps remain \emph{unchanged} for $D=4$ compared to that of $D=2$.
Since $D=4$ is the maximum possible ancilla dimension for both $N=2$ and $N=3$ cases~\cite{vidal_rank}, we conclude that the reported values of the gaps are \emph{intrinsic} to the gates, irrespective of the ancilla dimension.

\begin{table}[t]
\caption[Frobenius fidelity]{\label{table:F_D_2}The values of the Frobenius fidelity for sequential implementation of various global gates with an ancilla of dimension $D=2$.}
\centering
\vspace{3mm}
\begin{ruledtabular}
\begin{tabular}{c c c c c c c c }
Unitaries: &CNOT &CZ &CPHASE & SWAP & Toffoli & Fredkin  \\ [.5ex]
\hline
Fidelity & 0.7071 &  0.7071 & 0.9239 & 0.50 & 0.75 & 0.75\\  [.5ex]
\end{tabular}
\end{ruledtabular}
\end{table}

We believe also that such intrinsic values of the gaps are closely related to the entanglement content of each unitary operation, the
figure of merit for this purpose being the Schmidt strength~\cite{Nielsen2003} (proportional to the sum of nonzero Schmidt numbers of an operator) as a measure of \emph{operator entanglement}~\cite{Zanardi2001,Wang2002,Wang2003}. Exemplifying values of Schmidt strengths are 0.5 and 0.75 for CNOT and SWAP, respectively. As expected, the larger the Schmidt strength of the gate is, the larger the corresponding fidelity gap is.

\section{Conclusions}
\label{sec:conclusions}

In conclusion, we have studied the problem of sequential implementation of entangling unitary operations within the various languages of
Kraus representation, operator-Schmidt decomposition and MPO representation. Exploiting such tools allowed us to revisit the previously proposed no-go theorem through a systematic and instructive scheme that reveals various mathematical aspects of the subject. The Kraus representation approach, in particular, enabled us to probe possible correlations between qubits and an ancilla after unitary evolution of the joint system and thereby to compare the ultimate action of sequentially prepared unitary to that of a global entangler. Employing MPO formalism, we also developed numerical techniques for an efficient realization of a sequential version of an entangling unitary operation with an optimal fidelity.
As a realistic scenario, the recipe could instruct an experimentalist how to obtain optimal two-qubit operations that lead to a sequential
version of the desired multiqubit entangler with the highest possible fidelity and without the need to perform any measurement.

We stress that these results and the proposed optimization protocols can be of wide potential applicability at the interface of condensed matter physics, quantum optics, and quantum information. In particular, they will be of direct relevance to any sequential physical setup like photonic qubits, superconducting qubits, or quantum dots. From a methodological point of view, the proposed numerical technique, the VMPO method, promises to be a computationally efficient tool for variational search into the Hilbert-Schmidt space of operators. We believe that the method may particularly
be utilized for characterizing the entangling (and disentangling) properties of multiqubit unitary operations.

We point out, finally, that the no-go theorem does not refrain from implementing an entangling unitary if (instead of targeting
the nonlocal unitary within a \emph{single} sweep of the itinerant ancilla along the qubits chain) we allow the ancilla to sweep back and forth through the qubit chain and thereby perform \emph{multiple} bipartite operations at various stages on each qubit. For instance, it can be easily verified that CNOT
can be implemented through \emph{two} rounds of ancilla operations each performed in a sequential manner.

\begin{acknowledgments}

I gratefully acknowledge stimulating discussions with Enrique Solano, Lucas Lamata, David P\'erez-Garc\'{\i}a, Andreas Weichselbaum,
Jan von Delft, Jens Eisert, Mehdi Yaghouti, and Farshad Ebrahimi. I particularly would like to thank Andreas Weichselbaum for constructive comments on the manuscript. This work was supported by the Spintronics RTN, the DFG (SFB 631, De-730/3-2) and DFG under SFB 689. Financial support
from the German Excellence Initiative via the Nanosystems Initiative Munich (NIM) is gratefully acknowledged. I am also grateful to
Universidad del Pa\'{\i}s Vasco for support and hospitality.

\end{acknowledgments}

\bibliography{SIU}

\begin{thebibliography}{35}
\bibitem [{sch()}]{sch1}%
  \BibitemOpen
  \href@noop {} {}\bibinfo {note} {We adopt this convention in the remainder of
  the paper too, since exploiting the orthogonality feature of Schmidt vectors
  (or operators) alone suffices for the purpose of our arguments.}\BibitemShut
  {Stop}%
\bibitem [{fro()}]{frob}%
  \BibitemOpen
  \href@noop {} {}\bibinfo {note} {The Frobenius or Hilbert-Schmidt norm of an
  operator is defined by ${\|O\|}_F=\sqrt{\mathrm{Tr}(O^{\dagger} \:
  O)}$.}\BibitemShut {Stop}%
\end{thebibliography}

\end{document}